# On the generation of the particles through spontaneous symmetry breaking.

Voicu Dolocan
Faculty of Physics, University of Bucharest, Bucharest, Romania

**Abstract.** In this paper we present the Weinberg-Salam-Glashow model of the electroweak interactions. With a specific choice of parameters can be obtained massive Z and $W^{\pm}$ bosons, while keeping the photon massless. These results are obtained by breaking the local gauge invariant $SU(2)_L \times U(1)_Y$ symmetry. The same Higgs doublet which generates $W^{\pm}$ and Z masses is also sufficient to give masses to the leptons.

## 1. State of the Art

The Weinberg-Salam-Glashow model of leptons is based on the introducing of two vector fields, one isospin triplet $A'_\mu$ ($\mu = 1,2,3$) and one singlet $B_\mu$, which should finally result as fields of the physical particles $W^+$, $W^-$, $Z^o$ and photon, through the symmetry breaking induced by the Higgs mechanism [1-6]. The bosons $W^+$, $W^-$ and $Z^o$, mediating the weak interaction, must be very massive. The leptonic fields have to be distinguished according to their helicity. The helicity is associated with the sign of the scalar product **σ.p**, where **σ** is the spin and **p** is the momentum of the lepton. Every fermion generation ($e, \mu, \tau$) contains two related left-handed (negative helicity) leptons. These form an "isospin" doublet of left-handed leptons. There are also right-handed (positive helicity) components of the charged massive leptons. A right-handed neutrino does not exist (at least in the framework of weak and electromagnetic interactions). Therefore left-handed leptons can be represented by doublets

$$\Psi_L = \frac{1-\gamma^5}{2}\Psi = \frac{1-\gamma^5}{2}\begin{pmatrix} \nu \\ e \end{pmatrix} = \begin{pmatrix} \nu_L \\ e_L \end{pmatrix} \qquad (1.1)$$

while, right-handed leptons can be represented by singlets

$$\Psi_R = \frac{1+\gamma^5}{2}\Psi = \frac{1+\gamma^5}{2}e = e_R \qquad (1.2)$$

where

$$\gamma^5 = \begin{pmatrix} -I & 0 \\ 0 & I \end{pmatrix}; \qquad I = \begin{pmatrix} 1 & 0 \\ 0 & 1 \end{pmatrix} \qquad (1.3)$$

There are also the following relations

$$\overline{\Psi}_L = \overline{\Psi}\frac{1+\gamma^5}{2} = (\overline{v}\ \overline{e})\frac{1+\gamma^5}{2} = (v_L\ e_L)$$

$$\overline{\Psi}_R = \overline{\Psi}\frac{1-\gamma^5}{2} = \overline{e}\frac{1-\gamma^5}{2} = \overline{e}_R \qquad (1.4)$$

$$\overline{\Psi}\Psi = \overline{\Psi}_L\Psi_R + \overline{\Psi}_R\Psi_L = \overline{e}_L e_R + \overline{e}_R e_L$$

Glashow proposed that the Gell-Mann-Nishijima relation for the electron charge Q should also be valid in the case of the weak interaction

$$Q = T_3 + \frac{1}{2}Y \qquad (1.5)$$

where $T_3$ is the quantum number of the third component of isospin $\hat{T}$ and Y is the quantum number of hypercharge $\hat{Y}$. Since $\hat{T}_3$ and $\hat{Y}$ commute, both can be diagonal simultaneously. So in Eq. (1.5) we replace $\hat{T}_3$ and $\hat{Y}$ by their eigenvalues. For the known charge of the leptons (Q = -1) and neutrino (Q = 0) and from their classification with respect to isodoublets and isosinglets we can directly determine the $T_3$ and Y values of the various particles, as shown in Table 1.1.

Table 1.1. Weak isospin and hypercharge quantum numbers of leptons

μ 1,2,3

| Lepton | T | $T_3$ | Q | Y |
|---|---|---|---|---|
| $v_e$ | ½ | ½ | 0 | -1 |
| $e_L$ | ½ | -1/2 | -1 | -1 |
| $e_R$ | 0 | 0 | -1 | -2 |

The Lagrangian for the electron-neutrino lepton pair, which is invariant at SU(2)×U(1)$_Y$ gauge, is

$$L_1 = (\overline{v}_L\ \overline{e}_L)\gamma^\mu\left[i\hbar c\partial_\mu - g\frac{1}{2}\hat{\tau}\cdot\mathbf{W}_\mu - g'\left(-\frac{1}{2}\right)B_\mu\right]\binom{v_L}{e_L} +$$

$$\overline{e}_R\gamma^\mu\left[i\hbar c\partial_\mu - g'(-1)B_\mu\right] - \frac{1}{2}\mathbf{W}_{\mu\nu}\mathbf{W}^{\mu\nu} - \frac{1}{2}B_{\mu\nu}B^{\mu\nu} \qquad (1.6)$$

where was inserted hypercharge values $Y_L = -1$, $Y_R = -2$ and $\gamma^\mu = i\tau_\mu$ ($\mu = 1,2,3$). $L_1$ embodies the weak isospin and hypercharge interactions and final two terms are the kinetic energy and selfcoupling of the $W_\mu$ fields and the kinetic energy of the $B_\mu$ field

$$\mathbf{W}_{\mu\nu} = \partial_\mu \mathbf{W}_\nu - \partial_\nu \mathbf{W}_\mu - g\mathbf{W}_\mu \times \mathbf{W}_\nu$$
$$B_{\mu\nu} = \partial_\mu B_\nu - \partial_\kappa B_\mu \qquad (1.7)$$

We note that the left-handed fermion forms an isospin doublet, which transforms under $SU(2) \times U(1)_Y$ as follows

$$\begin{pmatrix} \nu_L \\ e_L \end{pmatrix} \rightarrow \begin{pmatrix} \nu_L \\ e_L \end{pmatrix}' = \begin{pmatrix} \nu_L \\ e_L \end{pmatrix} \exp\left[\frac{ig}{\hbar c}\mathbf{x}.\mathbf{W}.\hat{\mathbf{T}} + \frac{ig'}{2\hbar c}YBx\right] \qquad (1.8)$$

where g and g' are coupling constants and $\hat{T} = \hat{\tau}/2$. Under an infinitesimal gauge transformation

$$\begin{pmatrix} \nu_L \\ e_L \end{pmatrix}' = \left[1 + \frac{ig}{\hbar c}\mathbf{x}.\mathbf{W}.\hat{T} + \frac{ig'}{2\hbar c}\hat{Y}Bx\right]\begin{pmatrix} \nu_L \\ e_L \end{pmatrix} \qquad (1.9)$$

Therefore, in the Lagrangian we have replaced $\partial_\mu$ by the covariant derivative

$$D_\mu = \partial_\mu + \frac{ig}{\hbar c}\hat{\tau}_\mu \partial_\mu + \frac{ig'}{2\hbar c}\hat{Y}B_\mu \qquad (1.10)$$

Analogous

$$e'_R = \left[1 + \frac{ig'}{2\hbar c}\hat{Y}Bx\right]e_R \qquad (1.11)$$

and

$$D_\mu = \partial_\mu + \frac{ig'}{2\hbar c}\hat{Y}B_\mu \qquad (1.12)$$

$L_1$ describes massless bosons and massless fermions. Mass terms such as $(1/2)M^2 B_\mu B^\mu$ and $-mc^2 \overline{\Psi}\Psi$ are not gauge invariant and so cannot be added. The electron mass term may be written as

$$-mc^2\bar{e}e = -mc^2\bar{e}\left[\frac{1}{2}(1-\gamma^5)+\frac{1}{2}(1+\gamma^5)\right]e = mc^2\left[\bar{e}\frac{1}{2}(1-\gamma^5)e+\bar{e}\frac{1}{2}(1+\gamma^5)e\right] =$$

$$-mc^2\left[\bar{e}\left(\frac{1-\gamma^5}{2}\right)^2 e+\bar{e}\left(\frac{1+\gamma^5}{2}\right)^2 e\right] = -mc^2\left[\left(\bar{e}\frac{1-\gamma^5}{2}\right)\left(\frac{1-\gamma^5}{2}e\right)+\left(\bar{e}\frac{1+\gamma^5}{2}\right)\left(\frac{1+\gamma^5}{2}e\right)\right]$$

$$= -mc^2(\bar{e}_R e_L + \bar{e}_L e_R)$$

(1.13)

where we have used that

$$\left(\frac{1-\gamma^5}{2}\right)^2 = \frac{1-\gamma^5}{2}; \qquad \left(\frac{1+\gamma^5}{2}\right)^2 = \frac{1+\gamma^5}{2}$$

To generate the particle masses in a gauge invariant way we must use the Higgs mechanism. It was formulate the Higgs mechanism, so that $W^+, W^-$ and $Z^o$ become massive and the photon remains massless. To do this it is introduced a four real scalar field $\Phi$ and add to $L_1$ an SU(2)×U(1) gauge invariant Lagrangian for the scalar fields

$$\frac{2m}{\hbar^2}L_2 = \left|(i\partial_\mu - \frac{g}{\hbar c}\hat{\mathbf{T}}\cdot\mathbf{W}_\sigma - \frac{g'}{2\hbar c}YB_\mu)\Phi\right|^2 + V(\Phi)$$

(1.14)

$$V(\Phi) = m^2(\Phi^*\Phi) - \lambda(\Phi^*\Phi)^2$$

with $m^2 > 0$ and $\lambda > 0$. This potential will break (spontaneously) the symmetry. To keep $L_2$ gauge invariant the $\Phi$ must belong to SU(2)×U(1) multiplets. We arrange four fileds in an isospin doublet with weak hypercharge Y = 1

$$\Phi = \begin{pmatrix}\Phi^+ \\ \Phi^o\end{pmatrix}; \qquad \Phi^+ = \frac{\Phi_1 + i\Phi_2}{\sqrt{2}}; \qquad \Phi^o = \frac{\Phi_3 + i\Phi_4}{\sqrt{2}}$$

(1.15)

The potential V(Φ) of (1.14) has its minimum at a finite value of |Φ| where

$$\Phi^*\Phi = \frac{1}{2}(\Phi_1^2 + \Phi_2^2 + \Phi_3^2 + \Phi_4^2) = \frac{m^2}{2\lambda}$$

The manyfold of points at which V(Φ) is minimized is invariant at SU(2) transformation. We must expand V(Φ) about a particular minimum. The vacuum we choose has

$$\Phi_1 = \Phi_2 = \Phi_4 = 0; \quad \Phi_3^2 = \frac{m^2}{\lambda} = v^2 \qquad (1.16)$$

The effect is equivalent to the spontaneous breaking of the SU(2) symmetry. We now expand $\Phi(x)$ about the particular vacuum

$$\Phi_o = \frac{1}{\sqrt{2}} \begin{pmatrix} 0 \\ v \end{pmatrix} \qquad (1.17)$$

The result is that, due to gauge invariance, we can simply substitute the expansion

$$\Phi(x) = \frac{1}{\sqrt{2}} \begin{pmatrix} 0 \\ v + H(x) \end{pmatrix} \qquad (1.18)$$

into the Lagrangian (1.14). This vacuum, as defined above, is neutral since T = 1/2, $T_3$ = -1/2 and with our choice of Y = + 1 we have Q = $T_3$ + Y/2 = 0. This choice of the vacuum breaks $SU(2)_L \times U(1)_Y$ but leaves $U(1)_{EM}$ invariant, leaving only the photon massless. The gauge boson masses are identified by substituting thr vacuum expectation value $\Phi_o$ for $\Phi(x)$ in the Lagrangian $L_2$. The relevant term in (1.14) is

$$\begin{pmatrix} 0 & \frac{v}{\sqrt{2}} \end{pmatrix} \frac{1}{\hbar^2 c^2} \left| g \frac{\tau_1}{2} W_1 + g \frac{\tau_2}{2} W_2 + g \frac{\tau_3}{2} W_o + \frac{g'}{2} B \right|^2 \begin{pmatrix} 0 \\ v/\sqrt{2} \end{pmatrix} = \\ \frac{v^2}{4} \left[ g^2 W^* W^- + \frac{1}{2}(-gW_o + g'B)^2 \right] \frac{1}{\hbar^2 c^2} \qquad (1.18)$$

where we have used the following relations

$$W^{\pm} = \frac{1}{\sqrt{2}}(W_1 \mp iW_2)$$

$$\frac{1}{2}(\tau_1 W_1 + \tau_2 W_2) = \frac{1}{\sqrt{2}}(\tau^+ W^+ + \tau^- W^-) \qquad (1.20)$$

$$g^2(W_1^2 + W_2^2) = g^3(W^{+2} + W^{-2}) \qquad \text{or alternatively } 2g^2 W^+ W^-$$

$$\tau^+ = \frac{1}{2}(\tau_1 + i\tau_2) = \begin{pmatrix} 0 & 1 \\ 0 & 0 \end{pmatrix}; \qquad \tau^- = \frac{1}{2}(\tau_1 - i\tau_2) = \begin{pmatrix} 0 & 0 \\ 1 & 0 \end{pmatrix}$$

Comparing the first term (1.1) with the mass term expected for a charged boson $M_w^2 c^4$, we have

$$M_W = \frac{1}{2} v g W \qquad (1.21)$$

The second term of (1.19) is

$$\frac{v^2}{8\hbar^2 c^2}\left[g^2 W_o^2 - 2 g g' W_o B + g'^2 B^2\right] = \frac{v^2}{8\hbar^2 c^2}\left[g W_o - g' B\right]^2 + 0.\left[g W_o + g' B\right]^2 =$$

$$\frac{v^2}{8\hbar^2 c^2}(g^2 + g'^2) Z_o^2 + 0.A^2 = \frac{1}{\hbar^2 c^2}\left(\frac{1}{2} M_Z^2 c^4 + \frac{1}{2} M_A^2 c^4\right) \qquad (1.22)$$

The physical fileds $Z_o$ and A are defined by

$$Z_o = \frac{g W_o - g' B}{\sqrt{g'^2 + g^2}}; \qquad M_Z c^2 = \frac{v}{2}\sqrt{g'^2 + g^2}\, Z_o$$

$$A = \frac{g' W_o + g B}{\sqrt{g'^2 + g^2}}; \qquad M_A = 0 \qquad (1.23)$$

Denoting by

$$\frac{g'}{g} = \cos\theta \qquad (1.24)$$

may be rewritten

$$Z_o = -B\sin\theta + W_o \cos\theta$$
$$A = B\cos\theta + W_o \sin\theta \qquad (1.25)$$

and

$$\frac{M_W}{M_Z} = \cos\theta$$

$M_W$ is the mass of the charged bosons $W^\pm$ and $M_Z$ is the mass of the neutral $Z_o$ boson. Since the massless photon must couple with electromagnetic strength, $e$, the coupling constant define the weak mixing angle θ

$$e = g \sin \theta = g' \cos \theta \qquad (1.26)$$

The following relation is fulfilled

$$\frac{1}{2g^2 v^2 W^2} = \frac{1}{8 M_W^2 c^4} = \frac{G}{\sqrt{2}} \qquad (1.27)$$

where G is a universal constant with the empirical value G=1.136×10$^{-5}$ GeV$^{-2}$. One obtains gvW = 246 GeV and

$$M_W c^2 = \frac{37.3}{\sin \theta} \text{ GeV}, \qquad M_Z c^2 = \frac{74.6}{\sin 2\theta} \text{ GeV}$$

By studiing the momentum distribution of the emitted decay electrons and positrons the masses are measured to be

$$M_W c^2 = 81 \pm 2 \text{ GeV}; \qquad M_Z c^2 = 93 \pm 2 \text{ GeV}$$

which is in a good agreement with the predictions of the standard electroweak model. From the above relations may be determined sinθ. The electron mass term is not invariant under SU(2)$_L$×U(1)$_Y$ A term $\propto \bar{e}_L \Phi e_R$ is invariant under SU(2)$_L$×U(2)$_Y$. To genertae electron mass we modify the Lagrangian (1.12) as follows

$$L_e = -G_e \frac{1}{\sqrt{2}} \left[ (\bar{\nu}_L \; \bar{e}_L) \begin{pmatrix} 0 \\ v+H \end{pmatrix} \right] e_R + \bar{e}_R (0 \; v+H) \begin{pmatrix} \nu_L \\ e_L \end{pmatrix} =$$
$$= -\frac{G_e (v+H)}{\sqrt{2}} (\bar{e}_L e_R + \bar{e}_R e_L) = -\frac{G_e (v+H)}{\sqrt{2}} \bar{e} e = -\frac{G_e v}{\sqrt{2}} \bar{e} e - \frac{G_e H}{\sqrt{2}} \bar{e} e \qquad (1.28)$$

where the electron mass $mc^2 = G_e v / \sqrt{2}$. The last term is the electron-Higgs interaction. The mass of the electron is not predicted since G$_e$ is a free parameter. In that sense the Higgs mechanism does not say anything about the electron mass itself. The coupling of the Higgs boson to electrons is very small.

## 2. Conclusions.

We have presented the Weinberg-Salam-Glashow model of the electroweak interactions. With a specific choice of parameters can be obtained massive Z and W$^\pm$ bosons, while keeping the photon massless. These results are obtained by breaking the local gauge invariant SU(2)$_L$×U(1) symmetry. The same Higgs doublet which generates $W^{\pm}$ and Z masses is also sufficient to give masses to the leptons.